# Exploring ChatGPT and its impact on society

**Md. Asraful Haque**[1] · **Shuai Li**[2]

**Abstract**
Artificial intelligence has been around for a while, but suddenly it has received more attention than ever before. Thanks to innovations from companies like Google, Microsoft, Meta, and other major brands in technology. OpenAI, though, has triggered the button with its ground-breaking invention—"ChatGPT". ChatGPT is a Large Language Model (LLM) based on Transformer architecture that has the ability to generate human-like responses in a conversational context. It uses deep learning algorithms to generate natural language responses to input text. Its large number of parameters, contextual generation, and open-domain training make it a versatile and effective tool for a wide range of applications, from chatbots to customer service to language translation. It has the potential to revolutionize various industries and transform the way we interact with technology. However, the use of ChatGPT has also raised several concerns, including ethical, social, and employment challenges, which must be carefully considered to ensure the responsible use of this technology. The article provides an overview of ChatGPT, delving into its architecture and training process. It highlights the potential impacts of ChatGPT on the society. In this paper, we suggest some approaches involving technology, regulation, education, and ethics in an effort to maximize ChatGPT's benefits while minimizing its negative impacts. This study is expected to contribute to a greater understanding of ChatGPT and aid in predicting the potential changes it may bring about.

**Keywords** LLM · ChatGPT · Generative AI · RLHF · OpenAI

## 1 Introduction

The advancement of artificial intelligence has led to the development of several language models that have revolutionized the way people interact with machines [1–3]. One of the main benefits of large language models (LLMs) is their potential to learn the statistics of language. By processing large amounts of text data, these models are able to identify patterns and relationships between words and phrases, enabling them to accurately predict the next word in a sentence or classify the sentiment of a piece of text. This has made language models invaluable for tasks such as text completion, chatbot dialogue generation, and language translation. OpenAI's ChatGPT has gained major attention due to its ground-breaking features and applications [4–6]. Some of the specific tasks that ChatGPT can perform include generating human-like text, text summarization, language translation, storytelling, etc. [7, 8]. It can generate text that is grammatically correct, coherent, and contextually relevant. ChatGPT is rapidly becoming a ubiquitous presence in various facets of daily life, seamlessly integrating itself into diverse domains ranging from customer service to education to creative content generation. One of the most prominent examples is its utilization in customer service and support. ChatGPT is cost-effective, as it reduces the need for human agents to handle customer inquiries, saving businesses on staffing costs. Many companies are deploying ChatGPT-powered chatbots on their websites to provide quick and round-the-clock assistance to customers [9, 10]. ChatGPT improves customer service by providing a personalized experience, such as guiding users through troubleshooting processes and even initiating transactions. It automates tasks like scheduling appointments and answering frequently asked questions, saving time, and improving productivity. Creative content generation is yet another area

✉ Md. Asraful Haque
md_asraf@zhcet.ac.in

Shuai Li
shuai.li@oulu.fi

1 Computational Unit, Z.H. College of Engineering and Technology, Aligarh Muslim University, Aligarh, India

2 Faculty of Information Technology and Electrical Engineering, University of Oulu, Oulu, Finland

where ChatGPT is making its presence felt. Writers, artists, and content creators are leveraging the model's capabilities to generate ideas, craft compelling narratives, and even compose poetry. By providing a wealth of potential directions and creative angles, ChatGPT serves as a valuable source of inspiration and a catalyst for innovative thought processes. This infusion of AI-generated content into the creative landscape challenges traditional notions of authorship and offers new avenues for artistic expression. Furthermore, ChatGPT's integration extends to professional settings, where it aids in drafting emails, generating reports, and assisting in research tasks. Its versatility makes it a valuable asset for time-consuming tasks that involve generating text, as it can swiftly produce coherent and contextually relevant content.

However, while the integration of ChatGPT into various aspects of daily life brings undeniable benefits, it also raises important considerations [11–13]. ChatGPT can be prone to errors and glitches, which can impact the accuracy and reliability of the responses it generates. It may be biased towards certain types of language or topics, depending on the data that they were trained on [14]. The potential for biases and inaccuracies in AI-generated content demands a critical eye and careful evaluation. The risk of over-reliance on AI-generated responses underscores the importance of nurturing human critical thinking and analytical skills. Striking the right balance between AI assistance and human expertise is crucial to ensure that the technology complements human capabilities without overshadowing them. As ChatGPT continues to weave itself into the fabric of daily life, thoughtful implementation and ongoing refinement are essential to optimize its positive impact across different spheres.

The rest of this article is divided into four sections. Section 2 provides an overview of ChatGPT that includes the architecture of ChatGPT, its training process and the comparison of ChatGPT with other popular models. Section 3 explores the impacts of ChatGPT on several representative domains of the society. Section 4 suggests some changes that could be made to mitigate the potential misuse of ChatGPT. Finally, we conclude the paper in Sect. 5 by summarizing the key observations.

## 2 An overview of ChatGPT

### 2.1 Architecture

The Generative Pre-training (GPT) model, developed by OpenAI, represents a significant leap in natural language processing. GPT leverages a transformer-based architecture to learn language patterns from extensive textual data during a pre-training phase. This pre-training equips the model with a deep understanding of the statistics representing language structure and context. Vaswani A. et al. introduced this type of neural network in 2017 [15]. The success of GPT has led to the development of various versions, each building upon the previous one. GPT-1 pioneered the concept with 117 million parameters, demonstrating text generation capabilities [16]. GPT-2 expanded with 1.5 billion parameters and achieved more impressive text generation across diverse topics [17]. GPT-3 marked a milestone with up to 175 billion parameters, showcasing exceptional contextual understanding and versatility, handling tasks from translation to code generation [16, 18]. These versions collectively illustrate the evolution of generative AI, showcasing the tremendous progress made in text understanding, generation, and its far-reaching applications. The first version of ChatGPT was built on GPT-3. The next improved version of ChatGPT is based on the GPT-3.5 (Generative Pre-trained Transformer 3.5) model. The architecture consists of multiple layers of attention mechanisms and feedforward neural networks. Here is a textual representation of the key components of the GPT-3.5 architecture [18, 19]:

- Input embedding layer: This layer takes the input text and converts it into a numerical representation suitable for processing. This typically involves tokenization and embedding of the input text.
- Transformer encoder: GPT-3.5 uses a stack of transformer encoder layers. Each encoder layer has two main components: multi-head self-attention mechanism and feedforward neural networks. The attention mechanism allows the model to weigh the importance of different words or tokens in the input text while processing each word/token. After attention mechanisms, the model applies feedforward neural networks to process the information gathered from the attention layer.
- Normalization layer: A normalization layer is applied to the output of the feedforward layer to help stabilize the training process and prevent overfitting. The output of each layer is added to the input to create a residual connection, which helps to improve gradient flow and prevent vanishing gradients.
- Prediction layer: This is the output embedding layer. The final output of the model is generated by a linear transformation applied to the output of the last layer, followed by a softmax activation function.

The GPT-3.5 architecture also includes several advanced features, such as cross-lingual transfer learning and few-shot learning, which enable the model to transfer knowledge learned from one language to another and quickly adapt to new tasks with very little training data. The latest version of ChatGPT has been launched on GPT-4 model. As of debut, GPT-4 has not been able to accept images as input, although this capability is still possible [20, 21]. No information regarding GPT-4's technical specifications, including the model's exact size, has been provided by OpenAI [22]. However, the total number of parameters of GPT-4 is expected to be around 1.76 trillion [23].

## 2.2 Training process

The training process of ChatGPT involves mainly three stages —— pre-training, fine-tuning, and reinforcement Learning with Human Feedback [18, 24–27].

- Pre-training: The process begins by collecting a vast amount of text data from the Internet. This corpus of text contains a wide range of content from websites, books, articles, and other textual sources. The text data are often in multiple languages and covers various topics and domains. The collected text data are then tokenized, breaking it down into smaller units called tokens. Tokens can be as short as a single character or as long as a word or phrase. Tokenization helps in processing and representing the text for the model [25].The pre-training process of ChatGPT is primarily an unsupervised learning process. Unsupervised learning means that the model does not require explicit human-provided labels or annotations and relies on vast amounts of unlabeled text data to learn language understanding and generation capabilities. During pre-training, the model learns to predict the next token in a sequence of tokens. It takes a context window of tokens as input and predicts the probability distribution over the possible next tokens. This process helps the model learn the statistics of the grammar, syntax, semantics, and some world knowledge. Importantly, the model's capabilities for few-shot learning and cross-lingual learning are gradually developed during this phase. The model is exposed to a broad range of multilingual text data and a diverse set of tasks, which contributes to its few-shot learning and cross-lingual capabilities. However, these capabilities are refined further during fine-tuning.
- Fine-tuning: In this phase, the model is trained on certain tasks related to the user's need. The model is fine-tuned on a specific dataset that is generated with human annotators. For example, if ChatGPT is intended for customer support, it might be fine-tuned on a dataset containing customer queries and their corresponding responses. Human annotators provide examples of inputs and desired outputs (text prompts and desired responses). The model is then trained to predict the desired responses given the prompts. This is a form of supervised learning, because it uses labeled data with clear input–output pairs. Few-shot learning is facilitated by providing the model with a limited number of task-specific examples, enabling it to quickly adapt to new tasks with minimal data. Cross-lingual learning is further honed during fine-tuning as the model is exposed to multilingual text data. Fine-tuning optimizes the model for specific objectives, and thorough validation and testing ensure its suitability for various applications [26].
- Reinforcement Learning with Human Feedback (RLHF): After the supervised fine-tuning, reinforcement learning can be used to further improve the model's performance [27]. In this phase, the model interacts with an environment (e.g., a simulated conversation or a dialogue generation task) and receives rewards or penalties based on the quality of its responses. Human generated feedbacks, in the form of ratings, rankings, or corrections, serves as a reward signal for the model. That is why, it is called RLHF. Human feedback is used to create a reward function, guiding the model to align its behavior with human preferences. The process is iterative, with multiple rounds of fine-tuning and evaluation using human feedback. Once the model's performance meets desired standards, it can be deployed in real-world applications, where it provides more accurate and contextually relevant responses or actions.

Due to the thorough training approach (Fig. 1) that combines the advantages of pre-training, fine-tuning and RLHF, ChatGPT ultimately proves to be a versatile language model with the ability to handle a wide range of language-related tasks, opening the door for its use in various linguistic contexts and applications.

## 2.3 Comparison with other language models

Currently LLMs are mostly developed on the Transformer architecture, where multi-head attention layers are stacked in a very deep neural network [15, 27]. They are designed for natural language understanding and generation tasks. Training and deploying LLMs require significant computational resources, such as powerful GPUs or TPUs and extensive memory capacity, making them accessible mainly to large organizations and research institutions [27]. LLMs are characterized by their enormous number of parameters. Advances in LLMs have been driven by increasing model size. Researchers are continually exploring larger models to achieve even better performance on complex language tasks. ChatGPT is now considered as the largest, finest, and the fastest growing model currently in existence [9, 13]. This model has been trained on a massive corpus of diverse texts to learn the patterns and relationships between different words and phrases. It is able to generate coherent and fluent text that is often difficult to distinguish from text written by humans. Previous language models were not capable of achieving this kind of performance and accuracy. Earlier models were not as well suited for tasks that require generating human-like text or engaging in human-like conversations. Some specific features that make ChatGPT better suited are autoregressive, generative transformer architecture, large number of parameters, RLHF etc. Table 1 compares ChatGPT to four other language models based on following seven different features:

  (i)  Developer and release date
  (ii)  Model type: It refers to the underlying architecture of the LLM.
 (iii)  Number of Parameters: The number of parameters in a language model refers to the weights and biases that the model learns during training. Larger models typically have more parameters. For example, OpenAI's GPT-3 is known for having 175 billion parameters, making it one of the largest language models as of my knowledge cutoff in January 2022.
 (iv)  Dimension: The dimension usually refers to the size of the embedding space or the hidden layers in the model. In the context of a language model, this could be the dimensionality of the word embeddings or the hidden states in the model. Larger dimensions often allow the model to capture more complex patterns and relationships in the data.
  (v)  Learning rate: Learning rate is a hyperparameter that determines the size of the steps taken during the optimization process. It influences how quickly the model adapts to the training data. The appropriate learning rate can vary depending on the specific model architecture and dataset. It is a crucial param-eter that needs to be tuned to ensure effective training.
 (vi)  Context understanding: Context understanding refers to the model's ability to comprehend and retain information from the input text over an extended context. In the context of language models, it involves understanding the relationships between words and phrases in a given passage of text. Larger models with more parameters and capacity often exhibit improved context understanding, allowing them to generate more coherent and contextually relevant responses.
 (vii)  Application scope.

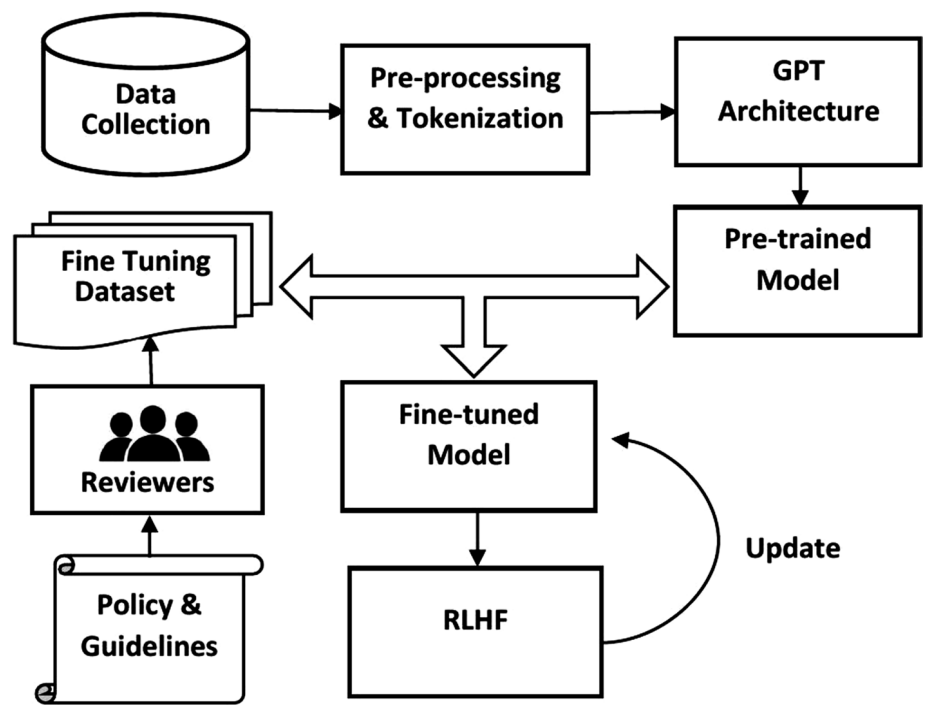

## 3 Impact of ChatGPT on society

According to Stephen Hawking [34], "…the rise of powerful AI will be either the best, or the worst thing, ever to happen to humanity. We do not yet know". The emergence of advanced artificial intelligence (AI) technologies has ushered in a new era of human–computer interaction, one characterized by the remarkable abilities of language models like ChatGPT. These models, built upon powerful transformer architectures and extensive training data, possess the capability to generate human-like text responses across a diverse range of subjects and contexts. As ChatGPT and similar language models become increasingly integrated into our daily lives, their impact on society is becoming more pronounced, influencing communication, education, creativity, and various other facets of modern life [35]. In this section, we will explore the multifaceted ways in which ChatGPT is shaping and reshaping the social landscape. The use of Chat-GPT has both benefits and risks [6, 34, 36]. ChatGPT can enhance the conversational experience between humans and machines, but it also has some negative aspects that need to be addressed.

**Table 1** Comparison among different language models

| Features | Language models | Description |
|---|---|---|
| Developer and release date [28–33] | ChatGPT | OpenAI (Nov-2022) |
| | BERT | Google (Oct-2018) |
| | T5 | Google (Oct-2019) |
| | GALACTICA | Meta AI (Nov-2022) |
| | LLaMA | Meta AI (Feb-2023) |
| Model type | ChatGPT | Autoregressive, Generative Pre-Trained Transformer |
| | BERT | Bidirectional Encoder Representations |
| | T5 | Text-to-Text Transformer |
| | GALACTICA | Language Translation or Analysis Model |
| | LLaMA | A multi-modal framework with the capability of Audio-Visual processing |
| Number of parameters [28–33] | ChatGPT | 175 billion parameters (GPT-3/3.5), 1.76 trillion parameters (GPT-4) |
| | BERT | 340 billion parameters |
| | T5 | 11-billion parameter |
| | GALACTICA | 120-billion-parameter |
| | LLaMA | 65 billion parameters |
| Dimension (L = number of layers; H = number of attention heads) [28–33] | ChatGPT | L = 96, H = 128 |
| | BERT | L = 24, H = 16 |
| | T5 | L = 24, H = 128 |
| | GALACTICA | L = 96, H = 80 |
| | LLaMA | L = 80, H = 64 |
| Learning rate [28–33] | ChatGPT | $6 \times 10^{-5}$ |
| | BERT | $5 \times e^{-5}$ |
| | T5 | $1 \times 10^{-2}$ |
| | GALACTICA | $7 \times 10^{-6}$ |
| | LLaMA | $1.5 \times 10^{-4}$ |
| Context understanding | ChatGPT | Unidirectional (Left Context) |
| | BERT | Bidirectional (Both Sides) |
| | T5 | Bidirectional Context |
| | GALACTICA | Bi/Multilingual Context |
| | LLaMA | Audio, image and text processing |
| Application scope | ChatGPT | Interactive text generation and dialogue |
| | BERT | Understanding context and tasks |
| | T5 | Text-to-Text Format conversion |
| | GALACTICA | Text translation and analysis |
| | LLaMA | Multimodal Context, audio processing |

### 3.1 Impact on education system

ChatGPT has the potential to impact the education system in both positive and negative ways. In the realm of education,

ChatGPT is emerging as a valuable tool for both students and educators. ChatGPT plays a significant role in assisting students with their education by offering a versatile learning companion that can provide valuable support across various academic domains. It serves as a readily available source of information, enabling students to swiftly access knowledge on a wide range of subjects, making it particularly useful for research and assignments. Additionally, it excels in simplifying complex concepts, helping students better understand challenging topics in subjects like science, commerce, and arts. Homework becomes more manageable with Chat-GPT's ability to offer explanations for problem-solving and essay writing. Language learners benefit from translation assistance, grammar corrections, and language practice exercises. For those pursuing computer science or programming, it can provide code explanations, debugging assistance, and coding challenges. Study tips, time management advice, and exam preparation strategies further aid students in achieving academic success. Moreover, ChatGPT fosters creativity by offering writing prompts, ideas, and feedback for creative writing endeavors, and it encourages critical thinking through discussions and debates. It can also introduce students to new topics, hobbies, or interests, promoting curiosity and self-directed learning. Accessibility is another key advantage as it provides information and support around the clock, catering to students' inquiries outside regular school hours. Ultimately, ChatGPT complements traditional learning methods, enhances access to information, and serves as a valuable educational resource across a broad spectrum of subjects and skills. Moreover, educators can use ChatGPT to create interactive learning materials, generate questions for quizzes and assessments, and facilitate personalized tutoring sessions. However, we cannot ignore the following negative impacts that are associated with the use of ChatGPT in education [37–42]:

- Reduced quality of learning: Overreliance on ChatGPT for learning could hinder critical thinking and deep understanding of subjects. Students might rely on generated answers without truly engaging with the material. The model's responses might not always be accurate or complete, leading to misconceptions and incomplete learning [43]. In general, ChatGPT tends to struggle with queries that can be easily addressed by humans due to limits in logical reasoning and background comprehension [44]. ChatGPT may greatly affect the problem-solving skills of the students. It might also lead to a decline in students' writing skills as they rely on the model to generate content rather than developing their own communication abilities.
- Decreased engagement with teachers: The use of ChatGPT may decrease students' engagement with teachers and their peers. Students may prefer to interact with the model rather than engaging in discussions and activities with their teachers and classmates, which could lead to a more isolated and passive learning experience. In subjects that involve interpersonal communication and human interaction, such as psychology and counseling, an overemphasis on AI-generated conversations might lead to a lack of empathy and understanding.
- Plagiarism and academic dishonesty: The ease with which ChatGPT can generate responses to questions and prompts could also lead to an increase in plagiarism and academic dishonesty [38, 43]. Students may be tempted to use the model to generate responses for assignments and assessments, rather than developing their own original work. Educators might face challenges in detecting AI-generated content in submissions.

## 3.2 Impact on job market

ChatGPT has the potential to disrupt the job market and increase unemployment in certain industries [45–48]. This is because it can perform many tasks that were previously done by humans. ChatGPT can summarize long texts into shorter, more digestible forms, making it easier for users to understand and process information. It can translate text from one language to another, making it possible for people who speak different languages to communicate with each other. ChatGPT can be used to generate content for websites, social media, and other platforms. This can include generating product descriptions, headlines, and social media posts. ChatGPT can be used to analyze the sentiment of a piece of text, which can be useful for understanding public opinion on a particular topic. ChatGPT can be used to automate customer service interactions, such as responding to frequently asked questions and handling customer complaints. It can also create engaging and imaginative stories with a coherent plot and characters. With such a broad range of abilities, ChatGPT is able to perform the job responsibilities mentioned in Table 2.

These are just a few examples of the many jobs that ChatGPT can perform. As the technology continues to develop, it is likely that it will be able to perform even more complex tasks related to natural language processing. It is clear that ChatGPT can be particularly useful for handling routine or repetitive inquiries, freeing up human representatives to focus on more complex or nuanced issues. Naturally, workers in customer service, data entry, and technical assistance may lose their jobs as a result of ChatGPT's automation. According to a report from Goldman Sachs [8, 56], 300 million full-time jobs could be impacted by AI systems like ChatGPT, escalating existing concerns around these platforms. Additionally, the use of ChatGPT can create a skills gap, where the demand for workers with advanced technical skills outstrips the supply. This can further exacerbate economic inequality, as those who have the skills to work with these technologies may be able to command higher salaries and better job prospects, while those who do not may be left behind. At the end, we want to mention that the deployment of ChatGPT can lead to the creation of some new job opportunities across industries [57]. These roles include AI trainers, ethicists, content moderators, developers, consultants, and administrators, among others.

## 3.3 Social isolation

While the technology has been widely lauded for its potential to revolutionize various industries, including healthcare,

**Table 2** The potential impact of ChatGPT on the job market

| Role | Job description |
| --- | --- |
| Customer support representative | ChatGPT can act as a virtual customer support representative, handling customer inquiries, troubleshooting issues, providing product information, and assisting with common problems. It can offer instant responses 24/7, enhancing customer satisfaction and reducing response times |
| Content creator | ChatGPT can assist content creators by generating blog posts, articles, social media captions, and marketing materials. It can help brainstorm ideas, provide outlines, and even suggest creative angles, streamlining the content creation process |
| Educational tutor | In the education sector, ChatGPT can function as an educational tutor. It can explain complex subjects, provide step-by-step explanations for problems, and engage in interactive learning conversations with students, enhancing understanding and retention |
| Translator | ChatGPT's language translation capabilities can be leveraged in translation services. It can quickly translate text from one language to another, offering accurate and contextually relevant translations for communication across diverse linguistic backgrounds |
| Creative consultant | For creative industries, ChatGPT can serve as a creative consultant. It can suggest plot twists for writers, generate story ideas, help artists brainstorm visual concepts, and even compose poetry or song lyrics |
| Data analyst | ChatGPT can assist in data analysis by summarizing large volumes of text data, extracting key insights, and generating reports based on textual information. It can aid in identifying trends and patterns within text-based datasets |
| Computer programmer | ChatGPT could assist with programming tasks. Given a set of requirements or specifications, it can produce code in a variety of programming languages [7, 49].ChatGPT can help diagnose errors in code [50, 51]. It can offer suggestions on where to look for bugs or how to approach debugging a particular issue |
| Legal assistant | In the legal field, ChatGPT can help draft legal documents, generate case summaries, and provide explanations of legal concepts. It can be used as a tool for legal research, analyzing case law, and offering initial insights into legal matters. The first AI legal assistant—"CoCounsel", which was launched in March 2023, is based on ChatGPT [52] |
| Medical assistant | ChatGPT can assist in medical settings by providing explanations of medical terminology, generating patient history reports, and simulating doctor-patient interactions for training purposes [53–55]. It can also offer general health information to patients |
| HR assistant | In human resources, ChatGPT can answer employee queries, provide information about company policies, and assist in drafting communication materials. It can streamline employee on boarding by answering common questions |
| Personal assistant | ChatGPT can function as a virtual personal assistant, helping users manage their schedules, set reminders, draft emails, and perform various administrative tasks |
| Marketing copywriter | ChatGPT can generate marketing copy, ad content, product descriptions, and promotional material. It can tailor messages to different target audiences and help marketers craft compelling narratives |
| Journalist and reporter | ChatGPT can assist journalists by generating news summaries, writing reports on factual events, and even suggesting potential angles for news stories |
| Financial analyst | In finance, ChatGPT can provide explanations of financial concepts, summarize financial news, and generate reports on market trends based on textual information |

finance, and customer service, there are also growing concerns about its social implications [58–60]. It may reduce human interaction. ChatGPT can be addictive, and people may become overly reliant on it for information or entertainment. While it can be useful for some purposes, it is not a substitute for human interaction. An over-reliance on ChatGPT for social or emotional needs can have detrimental effects on one's health, including social isolation and a decline in social contact and physical activity.

### 3.4 Social discrimination and bias

ChatGPT could generate responses that are racist, sexist, or homophobic. ChatGPT learns from large datasets on the internet, which may contain inherent biases and stereotypes present in human-generated text. It lacks real-time context understanding and relies solely on the patterns learned during training. Certain social groups are underrepresented in the training data and the model may not adequately learn to handle diverse perspectives. Therefore, if the training data reflect social prejudices, the model can inadvertently learn and reproduce those biases in its responses. Additionally, biased or discriminatory inputs by the users may also influence ChatGPT to generate insensitive responses.

### 3.5 Spreading misinformation

ChatGPT can be used to generate false or misleading information, which can spread rapidly on social media and other platforms [61–63]. This could have major repercussions, such as manipulating public opinion and damaging public confidence in information sources. It can create confusion, incite panic, and impact on decision-making. The spread of false information may contribute to social discord and distrust by creating divisions among people who believe different narratives.

## 3.6 Privacy and security issues

ChatGPT can potentially be used for unethical activities. It poses several security threats [64, 65]. Some examples are shown in Table 3.

## 3.7 Impact on environment

The environmental impact of large language models (LLMs) like ChatGPT can indeed be a concern, particularly in terms of CO2 emissions. Two main stages contribute to the carbon footprint: pre-training and user consultations.

- Pre-training:

  (i) Energy consumption: The pre-training phase of LLMs involves training on massive datasets for extended periods. This process requires significant computational resources and energy. The reported electricity bills for training large models like GPT-3 have indeed been substantial, often running into millions of dollars.

  (ii) CO2 emissions: The electricity used in pre-training, especially if derived from non-renewable sources, contributes to carbon emissions. The environmental impact depends on the energy mix of the data centers used for computation.

- User consults:

  (i) Inference and server costs: When users interact with LLMs, there are server costs associated with processing and generating responses. This involves running computations on powerful GPUs. The more users and queries, the higher the demand for server resources, contributing to increased energy consumption.

  (ii) Data center operations: The operation of data centers, which host the servers running LLMs, also requires significant energy. The cooling systems and other infrastructure contribute to the overall environmental impact.

It's important to note that these unethical activities are not inherent to ChatGPT itself, but rather the way it is used by individuals or organizations. To prevent these activities from happening, it is essential to ensure that ChatGPT is used responsibly and ethically, and that appropriate safeguards are in place to prevent abuse. Data minimization should be a priority, meaning that only essential information should be collected and shared with the AI model. Secure data transmission protocols like HTTPS should be utilized to protect data during communication. Implementing user authentication ensures that only authorized individuals can access ChatGPT, and access controls should be put in place to restrict actions and permissions. To respect privacy, adhere to data retention policies and anonymize user data, especially during training. Additionally, explore privacy-preserving techniques like federated learning or homomorphic encryption to safeguard user data during AI model development and inference [66]. Regular security audits and testing are crucial to identify and address vulnerabilities, while ensuring that the hosting infrastructure is secure and up to date [67]. Developers should follow ethical guidelines such as those from organizations like the OECD and stay informed about industry standards and best practices. Users should also be educated about the AI's limitations and data usage, and their consent should be obtained for data collection. Transparency is important, and there should be established channels for users to report concerns and obtain redress. Finally, compliance with relevant data protection regulations is essential, as is considering third-party audits for external validation of privacy and security practices. This comprehensive approach helps protect user data and ensures ethical and secure AI interactions.

**Table 3** Privacy and security concerns of using ChatGPT

| Threat types | How? |
| --- | --- |
| Invasion of privacy | ChatGPT requires large amounts of data to train effectively, and these data may contain personal or sensitive information. ChatGPT may retain data about user interactions and input, which could be used to build up a detailed profile of individuals. This could be used for targeted advertising or other nefarious purposes. There is a risk that these data could be leaked or stolen, potentially exposing sensitive information about individuals. ChatGPT could be used to generate messages that infringe on someone's privacy, such as by impersonating them in an online conversation |
| Increased risk of cyber-attacks | The use of ChatGPT could increase the risk of cyber-attacks, particularly if the model is not properly secured. Hackers may target the model to manipulate its responses for their own purposes. They can use the model to create text with links or dangerous code that could be used to steal personal information or install malware |
| Phishing and scamming | ChatGPT could be used to generate convincing messages that aim to trick people into revealing personal information, such as passwords or credit card details, or into performing actions that could harm them |
| Adversarial attacks | ChatGPT is vulnerable to adversarial attacks, in which an attacker tries to manipulate the output of the model by inputting carefully crafted text. This could be used to generate false information or cause the model to behave in unexpected ways |
| Cyberbullying and harassment | ChatGPT could be used to generate abusive or harassing messages that could be sent to individuals or groups, either anonymously or under false identities |

## 4 Mitigating negative impacts: some suggestions

Mitigating the negative impacts of ChatGPT and similar AI technologies involves a combination of technical, ethical, and policy-based measures. First and foremost, developers must prioritize transparency, ensuring that users are well informed about the capabilities and limitations of the AI system. In the context of ChatGPT, "hallucination" refers to the generation of responses that are factually incorrect or fabricated. Hallucinations can occur due to various factors, including a lack of real-time data, overfitting to training data, ambiguity in context, biases in training data, and the absence of source attribution [39]. ChatGPT has limitations in accessing external knowledge beyond their training data. If a query requires information not present in the training data, the model may generate speculative or hallucinated responses. It does not have the ability to fact-check information it generates in real time. It may provide responses that sound plausible but are not verified for accuracy. These inaccuracies can pose challenges in tasks requiring reliable information. Numerous studies have demonstrated that users must be cautious about the significant risks associated with relying on it [63, 68]. To address this issue, ongoing research aims to fine-tune models on domain-specific data, implement safety measures, and explore reinforcement learning from human feedback (RLHF).Ethical guidelines should be established to discourage harmful use cases and encourage responsible AI development. Robust data privacy and security measures must safeguard user information, and content moderation systems should filter out inappropriate or harmful content. Continuous monitoring for biases and misinformation is essential, with mechanisms in place to correct inaccuracies when detected. Monitoring and preventing the potential misuse of ChatGPT, particularly in relation to disinformation, propaganda, and cyberbullying, require a diverse strategy. Here are some strategies to consider:

### 4.1 Improved transparency and explainability

Enhancing transparency and explainability in AI model responses is essential for building trust and understanding in systems like ChatGPT. Various techniques and research directions can contribute to achieving these goals [69, 70]. For instance, attention mechanisms within models can be exposed to users, highlighting which parts of the input influenced the output. Adding interpretable layers to models can generate explanations alongside responses, providing users with insights into the decision-making process. Rule-based or template-based responses allow for predictability in specific contexts, while confidence scores offer users an indication of the model's certainty. Incorporating example-based or counterfactual explanations can help clarify responses further. Integrating external knowledge sources, such as knowledge graphs, can provide context and references. Users could also explicitly query the model for explanations, enabling a user-driven approach.

### 4.2 Content moderation and filtering

Content moderation and filtering is a crucial approach that entails deploying human moderators to actively review and assess the content generated by ChatGPT in real-time. Based on some guidelines, these moderators filter out or block content that contains offensive language, insults, or derogatory terms. Additionally, AI-based filtering systems can be developed and trained on a list of predefined keywords or phrases associated with disinformation, propaganda, hate speech, and cyber bullying [66, 67]. To prevent the spread of misinformation, it is essential to verify the content generated by ChatGPT; fact-checking tools can also be embedded [34]. These AI tools or tools can work alongside human moderators, automating content filtering processes and improving efficiency.

### 4.3 User reporting and feedback mechanism

Users should have the ability to report instances of misuse, disinformation, or cyberbullying directly within the application. These reports serve as triggers for a review process that can lead to content removal or appropriate user sanctions. Additionally, establishing a feedback loop between users and developers can be instrumental in refining the model's performance. User feedback can help identify emerging issues and guide continuous improvements [59].

### 4.4 Bias detection and mitigation

People are worried about biases in the design and impact of ChatGPT [61, 71]. To address biases in responses, especially those related to sensitive topics, bias detection and mitigation measures are vital, and that is why, OpenAI is currently researching "Rule based rewards" and "Constitutional AI" to make the fine-tuning process more understandable and controllable [71]. Our recommendation for managing biases is to conduct regular bias audits, especially external audits by enlisting independent researchers. Once these biases are identified, steps should be taken to reduce them and make sure the model produces fair and impartial results. Fairness assessment tools can also be employed to evaluate how ChatGPT interacts with various demographics and social groups, striving for equitable interactions.

### 4.5 Ethical guidelines and policies

Clear and comprehensive ethical AI guidelines and policies are essential for responsible usage. These guidelines should explicitly prohibit harmful activities, disinformation dissemination, and cyber bullying when interacting with ChatGPT [54, 58]. Another crucial part of developing ethical AI is making sure that the training data used for ChatGPT is properly selected to reduce biases and harmful material. A robust ethical framework sets the standards for acceptable usage and behavior.

### 4.6 Real-time updates and monitoring

To keep ChatGPT's responses contextually relevant and accurate, real-time updates and monitoring are necessary. One of the major drawbacks of ChatGPT is that it has limited knowledge of the world or events after 2021 [7]. It is because the system was trained with data available up to 2021. Integration of data feeds that provide real-time updates on current events and trends ensures that ChatGPT remains well informed and capable of providing up-to-date information. Continuous monitoring of user interactions and content generated by ChatGPT is equally crucial. Rapid response to emerging issues, coupled with adaptive adjustments to the model's behavior, helps maintain responsible usage.

### 4.7 User education

Educating the general public about the capabilities, limitations, and potential risks of ChatGPT is essential for responsible adoption and usage of these technologies. Public awareness campaigns can disseminate information through online content, videos, and educational materials. User-friendly interfaces should provide clear explanations of what AI systems can and cannot do. Educators can incorporate AI ethics and responsible use discussions into curricula, ensuring that future generations are well informed. Public demonstrations, workshops, and media coverage can offer insights into AI functioning and its implications. Moreover, platforms and forums for public engagement and community outreach can facilitate discussions and address concerns. Involving the public in decision-making processes related to AI, creating ethical AI certification labels, and establishing feedback mechanisms for users further promote transparency and awareness. Different organizations and research communities are working hard to educate, engage, and empower the public in the responsible use of AI technologies [72].

Mitigating the negative impacts of ChatGPT is an ongoing process that involves a collaborative effort from developers, policymakers, users, researchers, and various stakeholders. Developers must prioritize data ethics, transparency, and privacy by design, while also continuously monitoring and educating users about the technology's limitations and responsible usage. Policymakers play a crucial role in shaping the landscape by establishing regulations, ethical AI guidelines, and oversight mechanisms. They should also advocate for inclusivity in AI development teams and policymaking bodies to address potential biases. Users, on their part, need to stay informed and exercise control over their interactions with ChatGPT, utilizing available privacy settings and providing feedback to improve the technology. Researchers should actively work on reducing biases in AI models, share their findings transparently, and establish internal ethics committees to guide their work. Finally, stakeholders should form alliances and promote public engagement to facilitate discussions on AI ethics, ensuring diverse perspectives are considered. Efforts and initiatives are indeed underway to address the ethical, social, and employment concerns associated with ChatGPT. Organizations like OpenAI are actively involved in research, policy advocacy, and community engagement to improve responsible AI use. Academic institutions, think tanks, and industry groups are also contributing by conducting research and discussions on AI ethics and governance. In this collaborative ecosystem, researchers, developers, policymakers, and stakeholders collectively aim to create ethical AI standards and promote a responsible AI landscape that benefits society while minimizing potential harms.

## 5 Conclusion

ChatGPT has gained immense popularity in recent years, because it makes a lot of things easier and quicker. It is an evolved language model that has the potential to revolutionize the way we communicate with technology. It is an advanced natural language processing tool that has been trained on massive amounts of data, allowing it to generate human-like responses and understand complex language patterns. Its ability to understand and generate natural language has already made it a valuable tool in various industries, including customer service, content creation, and language translation. Students can turn to ChatGPT for explanations of complex subjects, seeking clarifications on difficult concepts in a conversational and accessible manner. ChatGPT holds immense promise, offering innovative solutions and transformative capabilities across various domains. However, it is vital to acknowledge that their integration into society brings forth both positive contributions and potential challenges. On one hand, ChatGPT enhances communication, streamlines tasks, and fosters accessibility, bolstering productivity and convenience. On the other hand, it raises concerns about biases, misinformation, privacy, and ethical dilemmas that demand careful consideration. To harness the benefits of ChatGPT responsibly, several key principles must guide its integration. Responsible usage entails transparency, ethical guidelines, and a commitment to bias mitigation. Regulatory frameworks should be embraced and adhered to, with audits and accountability measures to ensure ethical AI practices. Ethical considerations, such as respecting user privacy and fostering human–AI collaboration, should be at the forefront of AI development and deployment. Continuous improvement, both in the technology itself and in user education, is essential. We may use ChatGPT in a way that benefits society as a whole by being proactive in addressing these issues. Furthermore, the organization and developers should focus more on the promotion and presentation of ChatGPT as an AI tool to supplement rather than replace human capabilities. Promoting ChatGPT with a human-centric focus is vital for ensuring the responsible and ethical deployment of conversational AI technology. By placing humans at the center of the development and usage paradigm, it becomes possible to address user needs, concerns, and values effectively. Exploring the impact of ChatGPT is an ongoing and evolving process, and it requires a combination of technical, ethical, social, and policy-oriented approaches to address its multifaceted implications on society. The study of this paper is based on our understanding and knowledge gained from the review papers collected from different Internet sources. Therefore, it may struggle to cover the full spectrum of models, architectures, and applications. Another limitation of this study is that it may become outdated quickly due to new model releases, architectures, or advancements in understanding the strengths and limitations of the current version of ChatGPT. In future, we will try to address these issues through a systematic review.

**Author contributions** All authors equally contributed to the study conception, design and material preparation. The authors read and approved the final manuscript.

**Funding** The authors declare that no funds, grants, or other supports were received during the preparation of this manuscript.

## Declarations

**Conflict of interest** The authors have no relevant financial or non-financial interests to disclose.